%% file: main.tex
\documentclass[a4paper,fleqn,usenatbib,useAMS]{mnras}

\usepackage{graphicx}	
\usepackage{amsmath}	
\usepackage{amssymb}    
\usepackage{amssymb}	
\usepackage{multicol}	
\usepackage{bm}		
\usepackage{pdflscape}	
\usepackage{hyperref}
\usepackage{contour}
\usepackage{ulem}

\usepackage{xcolor}

\newcommand*\mean[1]{\bar{#1}}

\newcommand{\rockstar}{\textsc{Rockstar}}

\newcommand{\hmsun}{h^{-1}\ {\rm M_{\odot}}}

\newcommand{\hMpc}{h^{-1}\ {\rm Mpc}}

\newcommand{\ximm}{\xi_{\rm mm}}
\newcommand{\xihm}{\xi_{\rm hm}}

\newcommand{\rt}{r_{\rm t}}

\newcommand{\vt}{v_{\rm t}}
\newcommand{\Mt}{M_{\rm t}}
\newcommand{\Morb}{M_{\rm orb}}

\newcommand{\avg}[1]{\langle #1 \rangle}
\newcommand{\dsc}{\delta_{\rm sc}}

\newcommand{\deltasc}{\delta_{\rm sc}}

\newcommand{\aacc}{a_{\rm acc}}
\newcommand{\vr}{v_r}
\newcommand{\Mp}{M_{\rm p}}
\newcommand{\hMsun}{h^{-1}\ M_\odot}

\input{authorlist}

\title[A Better Halo Definition]{A Better Way to Define Dark Matter Haloes}

\begin{document}

\maketitle 

\label{firstpage}

\begin{abstract}
Dark matter haloes have long been recognized as one of the fundamental building blocks of large scale structure formation models.  Despite their importance --- or perhaps because of it! --- halo definitions continue to evolve towards more physically motivated criteria.  Here, we propose a new definition that is physically motivated, and effectively unique and parameter-free: ``A dark matter halo is comprised of the collection of particles orbiting in their own self-generated potential.''  This definition is enabled by the fact that, even with as few as $\approx 300$ particles per halo, nearly every particle in the vicinity of a halo can be uniquely classified as either orbiting or infalling based on its dynamical history. For brevity, we refer to haloes selected in this way as \it physical haloes. \rm  We demonstrate that: 1) the mass function of physical haloes is Press--Schechter, provided the critical threshold for collapse is allowed to vary slowly with peak height; and 2) the peak--background split prediction of the clustering amplitude of physical halos is statistically consistent with the simulation data, with an accuracy no worse than $\approx 5\%$.
\end{abstract}


\section{Introduction}
\label{intro}

Dark matter haloes are one of the basic building blocks for models of large scale structure \citep{Cooray-Sheth}.  Despite their importance, a unique, physics-based definition of dark matter haloes has remained elusive.  Typically, haloes are defined as spheres in which the mean matter density is some factor of a reference density, such as the critical density or the matter density of the Universe \citep[e.g.,][]{lacey_94}. The choice of this factor is arbitrary and varies in the literature, although values around 200 are commonly chosen based on the spherical collapse model \citep{gunn_72}. In addition, spherical overdensity definitions suffer from theoretical artifacts such as the need to characterize pseudo-evolution \citep{diemer2013a}, and practical issues such as subhalo stripping beyond the ``virial'' radius \citep{bahe_13, behroozi_14}.

Over the past several years, a number of studies have moved the field toward more physical definitions of dark matter haloes.  For example, it has long been recognized that gravitationally bound subhaloes in a halo extend beyond a halo's ``virial radius'' \citep[e.g.,][]{Baloghetal00,Mamonetal04, gill_05}. For this reason, the splashback radius, defined as the radius where particles reach the apocenter of their first orbit, has been proposed as a more physically motivated halo boundary \citep{diemerkravtsov14, adhikari2014,  More2015, Shi2016, shellfish}. Remarkably, haloes defined using the splashback radius exhibit a mass function that is much closer to universality than standard halo definitions \citep{Diemer20}.  However, the distribution of apocenters is broad \citep{diemer2017}, making it difficult to robustly define a single radius that encloses all orbiting particles or all satellites \citep{Diemer2021}.

A similar conclusion was reached by \citet{GarciaRozo2021}, albeit from an entirely different line of argument.  They proposed the transition between the one- and two-halo terms as a natural boundary, which was identified with the valley between the ``two humps'' observed in $r^2\xi(r)$, where $\xi$ is the halo--mass correlation function \citep[see also][]{fong_21}.  Doing so dramatically simplified the halo model while improving its accuracy.  Moreover, these improvements were achieved while self-consistently redefining subhaloes based on the newly proposed halo definition.  This is important since the choice of ``percolation'' (i.e. subhalo identification) impacts halo statistics \citep{GarciaRozo2019}.  The halo radius that emerged from this analysis was larger than the splashback radius that includes 90\% of particle apocenters.

In parallel to the these developments, there has been an increasing recognition that the dynamical structure of dark matter haloes can be used to motivate more physical halo definitions.  The foundational insight for our work is the fact that the phase space distribution of particles around haloes exhibits two distinct components. These components are: 1) particles orbiting in the halo potential; and 2) particles falling into the halo potential for the first time \citep[e.g.,][]{fillmore_84, Fukushige01, cuesta_08, diemand_08}. This dichotomy is particularly important for dynamical modeling of galaxies in haloes \citep[e.g.,][]{oman_13, ZW2013, adhikari_19, Hamabata2019}, and has been unambiguously detected in the Sloan Digital Sky Survey \citep{Tomooka2020}. However, it is only recently that authors have attempted to classify individual particles/galaxies in simulations into these two categories \citep{ sugiura_20, adhikari_21, bakels_21, Aung2021, aungetal22, diemer2022a, diemer2022b}.

Motivated by multiple lines of work that point towards a more physical definition of a halo, one that relies on its dynamics, we study the phase space structure of particles around haloes. As expected, the distribution in phase space is bimodal, although with significant overlap between the orbiting and infalling particle populations. 
We develop a method for robustly classifying particles into orbiting and infalling based on each particle's dynamical history.  In turn, this classification allows us to redefine haloes as the collection of all the orbiting particles in a structure. We then study the implications this new definition has on halo statistics, specifically halo abundance and clustering bias.

In the last few months, \citet{diemer2022a, diemer2022b} published a closely related work, arguing that one can conceptually differentiate between orbiting and infalling particles based on the first pericentric passage.  Their algorithm for identifying pericentres is superficially similar but quite different from ours in practice: our algorithm works in the $r$--$v_r$ subspace of the full 6D phase-space, while Diemer's works in 3D configuration space plus radial velocity.  Therefore, we emphasize that \it all our results concerning the pericentre definition are exclusive to the specific algorithm described in this paper \rm and need not apply to the algorithm of \citet{diemer2022a}. We do, however, expect the different orbiting/infalling classifications to agree for the vast majority of particle orbits. We further emphasize that the goals and methods of our work are both different and complementary.  Specifically, Diemer relies on standard halo definitions throughout and focuses on the characterization of the orbiting and infalling density profiles as a function of mass, accretion rate, redshift, and cosmology.  By contrast, our work relies on the orbiting/infalling split to argue for a fundamental redefinition of dark matter haloes, followed by a characterization of the resulting halo mass function and linear clustering bias.  

This paper is organized as follows. Section \ref{sec:sim-data} describes the simulation and halo catalogs we used. Section \ref{sec:Rafa-split} presents the classification algorithm to separate orbiting and infalling particles. Section \ref{sec:Edgar-split} presents an alternative but consistent definition of orbiting/infalling and how its results compare to those of the algorithm in section \ref{sec:Rafa-split}. Section \ref{sec:implications-lss} presents our newly proposed halo definition, and characterizes the resulting halo mass function and clustering bias.  We summarize our findings in section \ref{sec:summary}.

\section{Simulation and Halo Catalogs}
\label{sec:sim-data}

\begin{figure*}
    \includegraphics[width=\textwidth]{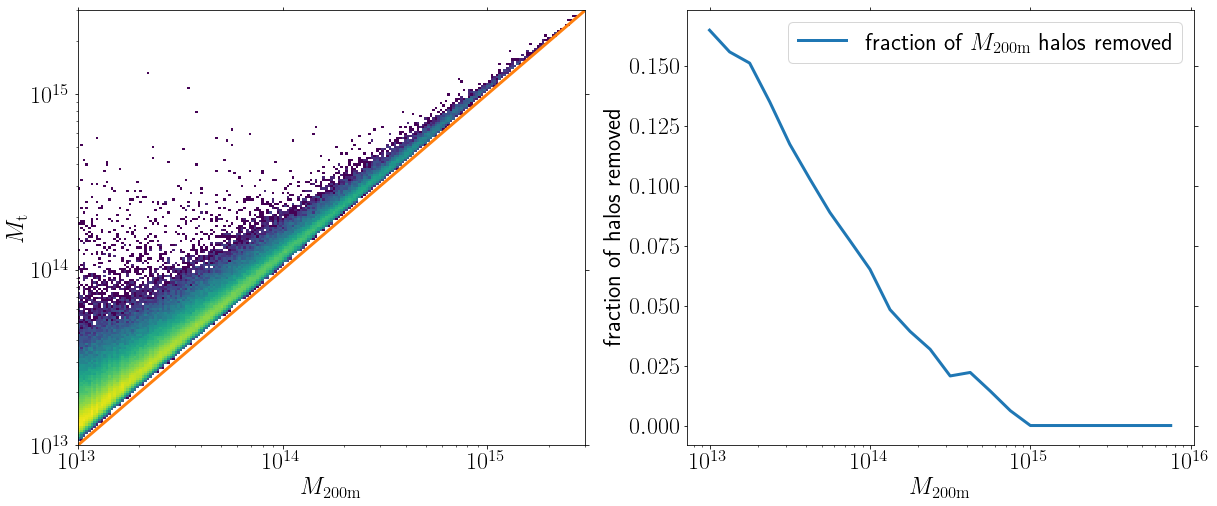}
    \caption{{\bf Left:} The relation between the mass $\Mt$ advocated in \citet{GarciaRozo2021} and the original halo mass $M_{200{\rm m}}$.  The orange line corresponds to $y=x$, and demonstrates that all haloes have $\Mt \geq M_{200{\rm m}}$. This inequality arises from the fact that the transition radius $\rt \geq r_{200}$ for all haloes. {\bf Right:} The fraction of \rockstar\ haloes that fall within the radius $\rt$ of a more massive halo, and are therefore removed from the $\Mt$-based halo catalog.}
    \label{fig:mt-m200m}
\end{figure*}

This work relies on a Cold Dark Matter (CDM) simulation run with Gadget2 \citep{Springel2005} that was first described in \citet{banerjeeetal20}.  The simulation box is $1\ h^{-1}{\rm Gpc}$ across, contains $1024^3$ particles, and has a softening length of $0.015\ \hMpc$.  The cosmological parameters are $\Omega_{\rm m}=0.3$, $\Omega_\Lambda=0.70$, $n_s=0.96$, $h=0.7$, and $A_s=2\times 10^{-9}$, corresponding to $\sigma_8=0.85$.  The initial conditions are set at $z=99$ using \texttt{N-GenIC} \citep{ngenic}.  An initial halo catalog obtained using a standard spherical overdensity definition with an overdensity threshold of 200 with respect to mean was constructed using the \rockstar\ halo finding algorithm \citep{Behroozi2013}.

In \citet{GarciaRozo2021} we argued that a simple yet accurate halo model of the halo--mass correlation function can be achieved if the radius $\rt$ at which the one-halo term is truncated is treated as a free parameter.   There, we found that the best fit radius $\rt$ was approximately equal to the radius corresponding to the minimum of $r^2\xihm(r)$.  For this work, we skip the model-fitting and set $\rt$ to the radius corresponding to this minimum. We use these results to refine the original \rockstar\ catalog as follows:

\begin{enumerate}
    \item Measure the halo--mass correlation function in bins of $M_{200{\rm m}}$, where the reference density is $\rho_{\rm m}$.
    \item Determine $\rt$ for each mass bin, and compute the corresponding integrated mass within $\rt$, which we label $\Mt$. 
    \item Fit the $\Mt$--$\rt$ relation as a power-law
    \item For each halo in the original halo catalog, we compute a new halo radius by the intersection of the integrated mass profile $M(r)$ with the $\Mt$--$\rt$ relation.  Note that if we were to replace the $\Mt$--$\rt$ relation by the $M_{200{\rm m}}$--$r_{200}$ relation this procedure would recover $M_{200{\rm m}}$.
    \item We percolate the resulting halo catalog by rank-ordering the haloes by $\Mt$.  Any haloes within $\rt$ of the most massive halo are removed from the halo catalog. We then move to the next most massive halo, and the procedure is iterated until we go through the full halo catalog.
\end{enumerate}

The end result is a halo catalog where the haloes have been self-consistently defined and percolated using the radius $\rt$ and the $\Mt$--$\rt$ relation recovered from the data. The left panel of Figure~\ref{fig:mt-m200m} shows the mass $\Mt$ of the haloes as a function of the more standard $M_{200{\rm m}}$ mass.  The right panel of Figure~\ref{fig:mt-m200m} shows the fraction of haloes as a function of $M_{200{\rm m}}$ that appear in the original catalog, but which are missing from our $\Mt$ catalog because they fall within the radius $\rt$ of a more massive halo.  It is this halo catalog that we use as the starting point for the analyses presented in this paper. We consistently measure quantities for stacked haloes binned by mass.  In units of $\log_{10} \Mt$ where $\Mt$ is $h^{-1} M_\odot$, the mass bin edges are: $\log_{10} \Mt$: $[$13.50, 13.60, 13.70, 13.85, 14.00, 14.30, 14.50, 14.60, 14.70, 14.85, 15.00, 15.50$]$. This corresponds (roughly) to haloes of mass $M_{200{\rm m}} \geq 2\times 10^{13}\ \hMsun$. The minimum halo mass is set by the requirement that haloes contain at least 300 particles.

\section{Establishing the Orbiting/Infall Dichotomy}
\label{sec:Rafa-split}

\begin{figure*}
    \includegraphics[width=\textwidth]{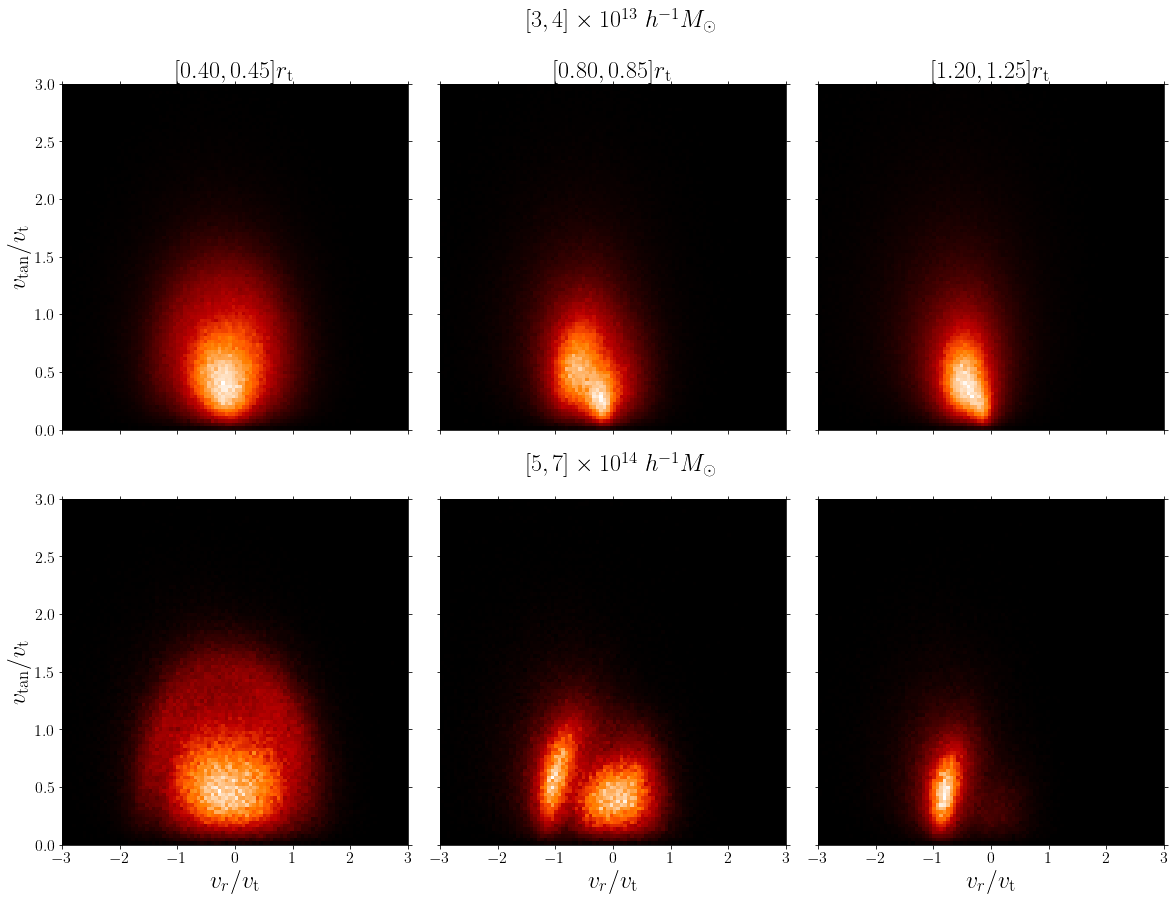}
    \caption{Two dimensional distributions of radial and tangential velocities of particles around haloes in two mass bins.  The color traces the density of points in velocity space in arbitrary units. The top row is for a mass bin $\Mt \in [3,4]\times 10^{13}\ \hMsun$, while the bottom row is for a mass bin $\Mt \in [5,7]\times 10^{14}\ \hMsun$. We show the distributions of radial and tangential velocities for three radial bins, as labelled. \textbf{Left:} The inner region consists mostly of orbiting particles with a small infalling component, which populates the high-velocity ``ring'' seen in high-mass haloes.  \textbf{Center:} Closer to the halo boundary the infalling population becomes more prominent and exhibits a large negative radial velocity. \textbf{Right:} Even past the nominal halo boundary $\rt$ there remains a small population of orbiting particles.}
    \label{fig:phase-space-1}
\end{figure*}

\begin{figure*}
    \includegraphics[width=\textwidth]{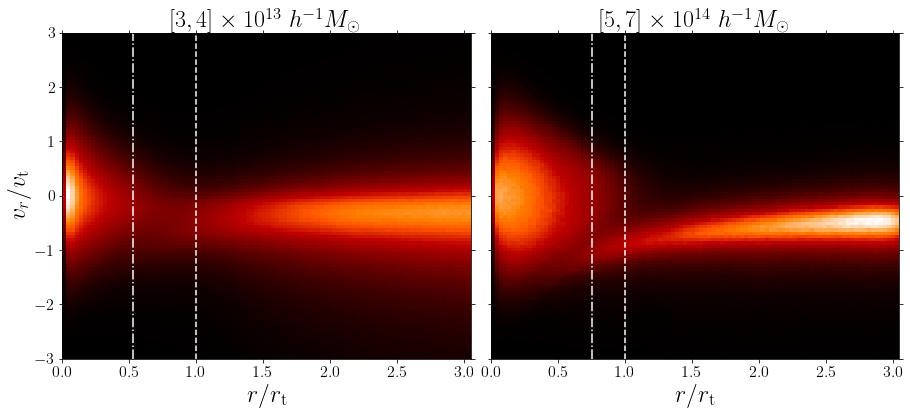}
    \caption{Two dimensional distribution of particles in the $r$-$v_r$ plane for two mass bins, as labelled. The particle distribution is obviously bimodal.  The ``triangle'' at low radii corresponds to orbiting particles, while the infall stream with negative radial velocities extending to large radii corresponds to infalling particles. The two particle populations overlap in real space.  Consequently, no spherical-halo definition can adequately separate these two components.  Shown for reference is the average radius $r_{200}/\rt$ (dot dashed) of the haloes, as well as the radius $\rt$ (dashed).}
    \label{fig:phase-space-2}
\end{figure*}

\begin{figure*}
    \includegraphics[width=\textwidth]{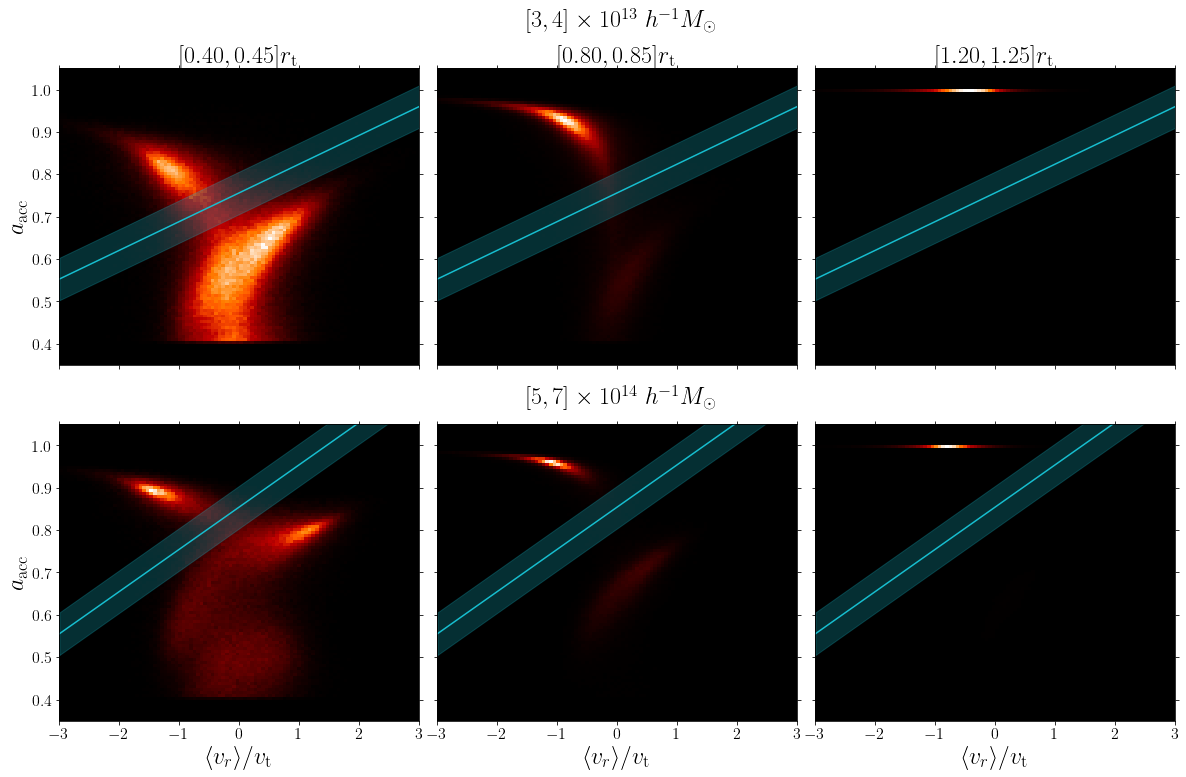}
    \caption{Two dimensional distribution of particles in the $\langle v_r \rangle$-$a_{\rm \aacc}$ plane. We show the distributions for three representative radial bins for the same low mass (top row) and high mass (bottom row) bins as in Figure~\ref{fig:phase-space-1}. The cyan line defines the cut we use to split the particles into their orbiting (bottom right part of the plot) and infalling (top left) populations. The width of the band corresponds to the volume used to define this split (see text for details).}
    \label{fig:Rafa-split}
\end{figure*}

Motivated by the findings of \cite{Aung2021}, we study the phase space structure of particles around dark matter haloes. We follow the conventions of \cite{Aung2021}. The radial velocity of a particle is defined as $v_r = \vec{v} \cdot \hat{r}$, where $\hat{r}$ is the radial unit vector from the center of a halo to a particle. The radial velocity is positive for outgoing particles. The tangential velocity is defined as $v_{\rm tan} = \sqrt{v^2 - v_r^2}$, which is always positive.

Figure \ref{fig:phase-space-1} shows the distribution of radial and tangential velocities of particles in spherical shells around dark matter haloes for a high mass bin ($\Mt \in [5, 7] \times 10^{14}\ \hMsun$) and low mass bin ($\Mt \in [3,4]\times 10^{13}\ \hMsun$) bin. We show this distribution for three representative radial bins that go from the inner regions of the haloes to past their radius $\rt$. We have rescaled velocities by $v_{\rm t} \equiv \sqrt{G \Mt / \rt}$. For the high mass haloes that were studied by \cite{Aung2021}, our results are similar to theirs: the distribution of particle velocities is clearly bimodal.  The cloud of particles scattered around $v_r\approx 0$ corresponds to orbiting particles, while the ring/arc of high velocity particles with $v_r<0$ are infalling particles.  Turning to the velocity structure of low mass haloes, we see that while the particle distribution appears bimodal, the two populations exhibit strong overlap.

We can gain additional insights by plotting the particle distribution in the $r$-$v_r$ plane. Figure~\ref{fig:phase-space-2} shows the distribution of particles in the $v_r$ vs $r$ space, typically referred to as a phase space plot. The velocities and radii are re-scaled by $\vt$ and $\rt$ respectively.  We see that the distribution of radial velocities at fixed radius is clearly bimodal across a broad range of radii.  The orbiting component corresponds to the cloud of points at small radii that scatter around $v_r=0$.  Note that this orbiting component extends out to $r/\rt\approx 1$ (vertical dashed line), whereas many orbiting particles lie beyond the radius $r_{200}/\rt$ (vertical dash-dot line).  The infall component is the stream of particles with negative radial velocities that extends out to large radii.  Note that these two particle populations can \it not \rm be cleanly separated with a cut in either Figure~\ref{fig:phase-space-1} or \ref{fig:phase-space-2}.  The overlap of the two particle populations demonstrates that \textit{no spherical halo definition can adequately separate orbiting from infalling particles.}

Given that we are unable to make a simple phase space cut that cleanly separates orbiting from infalling particles, we turn to the particle dynamical histories as a means of distinguishing between these two populations.  Specifically, we rely on two key expectations: 1) infalling particles will exhibit late accretion times; and 2) the average radial velocity of orbiting particles should be $\approx 0$.

We define the accretion time $\aacc$ of a particle as the value of the expansion factor $a$ when the particle crosses the halo radius $\rt$ for the first time.  The boundary $\rt$ used to define the accretion epoch is calculated once at $z=0$.  To calculate the radial separation, the location of the halo center as a function of time is taken to be the location of the halo progenitor in the main branch of the merger tree. The epoch $\aacc$ is calculated for all particles within $~10$ Mpc/h of each of our halo centers using 100 time snapshots between $a=0.2$ and $a=1$.  Turning to our use of radial velocities, we expect orbiting particles to have low mean radial velocities, whereas infall particles will have large negative radial velocities. We considered averaging the radial velocities over a range of times of the form $\lambda t_{\rm dyn}$ where $t_{\rm dyn} \equiv \sqrt{\rt^3 / G \Mt}$ is the halo dynamical time and $\lambda$ is of order unity. They all performed comparably well, though the orbiting/infall split was especially stable for $\lambda$ in the range $[1/8,1/2]$. This is a shorter interval than $\lambda=2$, corresponding to the crossing time for a particle that is initially at rest. However, infalling orbits have high initial radial velocities, which favor $\lambda < 1$ (i.e. they need less time to cross the halo).

Figure~\ref{fig:Rafa-split} shows the distribution of particles in the vicinity of a halo in this plane. The top and bottom rows correspond to low and high mass haloes respectively, while each column corresponds to a different radial bin.  We find that the particle distributions are bimodal, with each of the two components being largely disjoint from the other. Particles in the top-left part of the plot have late accretion times and negative radial velocities, and are therefore infalling particles.  Particles in the bottom right of the plot have early accretion times and velocities that scatter around $\avg{\vr}\approx 0$. The splashback population are orbiting particles with comparatively late accretion (relative to other orbiting particles) and large mean radial velocities.

Given that there is little overlap between the orbiting and infalling particle populations in this space, we distinguish between the two populations using a linear cut (cyan line in Figure~\ref{fig:Rafa-split}).  This line is chosen by minimizing the number of particles in a narrow band around the line.  Specifically, we define the cost function as the number of objects within a distance $\epsilon$ from the line, where ``distance'' is the ordinary Euclidean distance. We then find the slope and y-intercept of the line which minimizes this cost function.  We use $\epsilon = 0.05$, though we have explicitly verified that the split is robust to the choice of $\epsilon$. We also tested the robustness of this method to the number of time steps used when defining $\avg{v_r}$ and $a_{\rm acc}$ by using subsets of the total number snapshots in the simulation.  We find that our orbiting/infall decomposition is numerically robust. The details of these time resolution tests are included in Appendix~\ref{app:convergence}.

We caution that while the particle distribution in the $\aacc$--$\avg{v_r}$ space is strongly bimodal (Figure~\ref{fig:Rafa-split}), the two populations are still not completely disjoint. This is clearly seen when plotting the particle density with a logarithmic color scale reveals that the tails of the orbiting and infalling populations overlap somewhat, particularly at small radii (see Figure~\ref{fig:Rafa-split-log}). In the next section we provide some sense of the (small) theoretical uncertainty associated with this overlap.

\section{When Does a Particle Transition from Infalling to Orbiting?}
\label{sec:Edgar-split}

\begin{figure}
    \includegraphics[width=\linewidth]{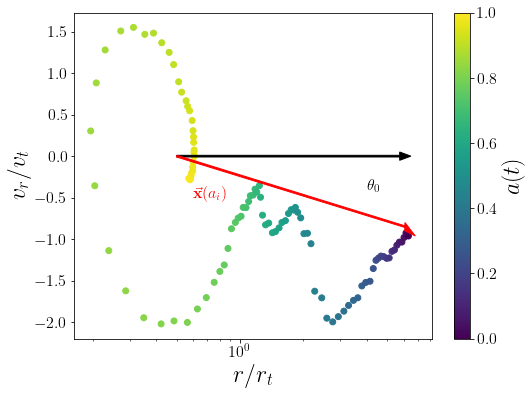}
    \caption{Diagram of a particle's orbit in phase space  as a function of time, indicated by the scale factor, $a(t)$, in color. The angle swept by the particle is measured from the black arrow. A particle reaches its first pericenter when the cumulative angle is $\pi$.}
    \label{fig:Edgar-split}
\end{figure}

\begin{figure*}
    \includegraphics[width=\textwidth]{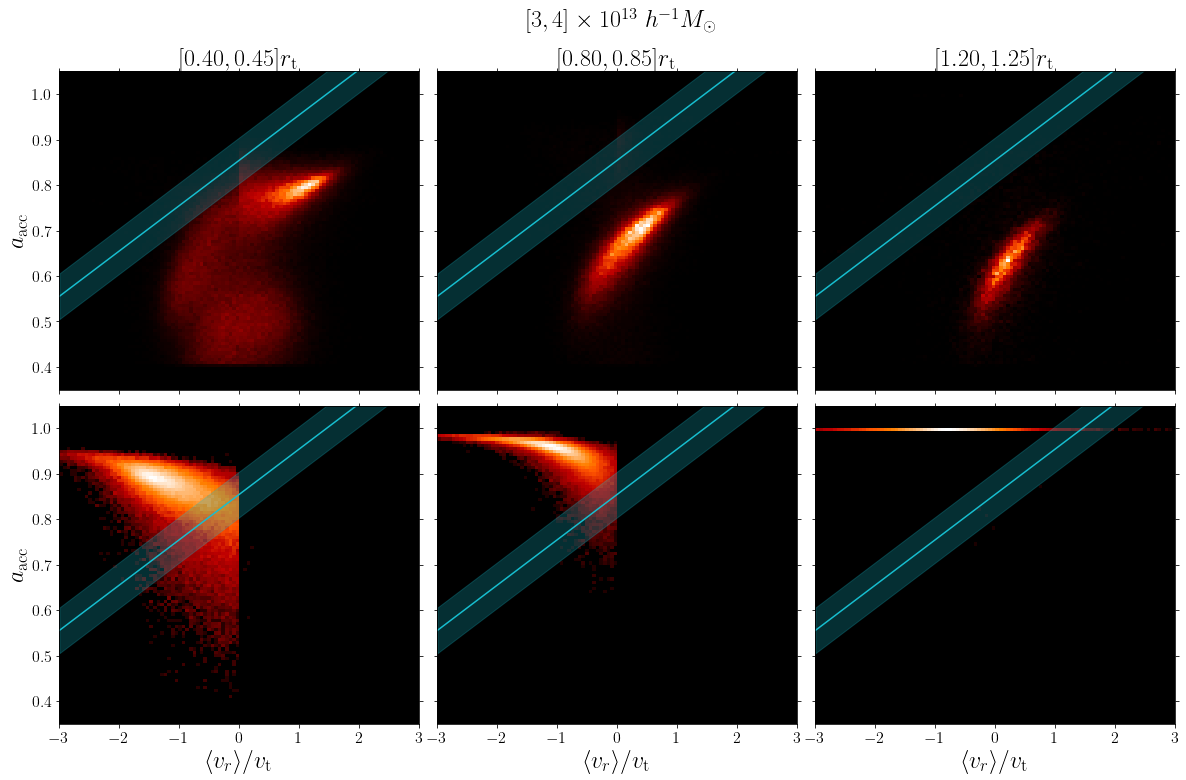}
    \caption{Distributions of orbiting (top row) and infalling (bottom row) particles classified by the pericenter method in the $\langle v_r \rangle$-$\aacc$ plane. The blue line is the classification boundary defined in section \ref{sec:Rafa-split}. The new pericenter classification is generally in good agreement with our accretion-based classification, which larger differences becoming apparent at small radii. Importantly, there is an obvious discontinuity in the radial velocities of orbiting and infalling particles as defined using our pericenter splits.  Consequently, some particles must be miss-classified in this alternative approach. The bottom left corner shows that, in the pericentric classification, there exists a population of infalling particles with early accretion times. Given that the obvious discontinuity implies some particles are miss-classified, it is likely that these early accretion infalling particles are better characterized as orbiting.}
    \label{fig:high-splits-comparison}
\end{figure*}

\begin{figure}
    \includegraphics[width=\linewidth]{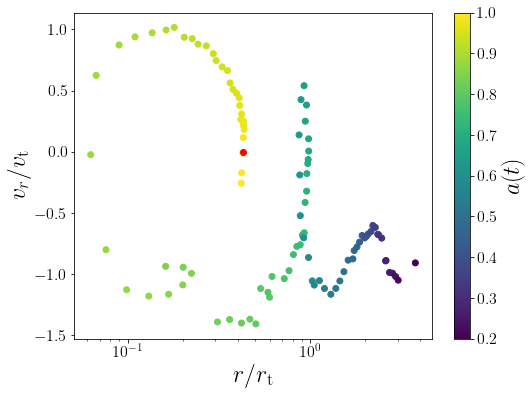}
    \caption{The phase-space orbit of a particle that is: 1) classified as infalling using our pericentric passage algorithm; and 2) has an early accretion time.  The orbits for the majority of particles satisfying these two criteria are qualitatively similar. Evidently, these particle should have been tagged as orbiting.  The red point marks the radius $r_{\rm c}$ that our algorithm selects for determining the ``angle swept'' (see Figure~\ref{fig:Edgar-split}).  The fact that this red point falls just shy of the zero crossing of the particle's orbit explains why this particle was miss-classified.}
    \label{fig:missclassified-orbit}
\end{figure}

\begin{figure*}
    \includegraphics[width=\textwidth]{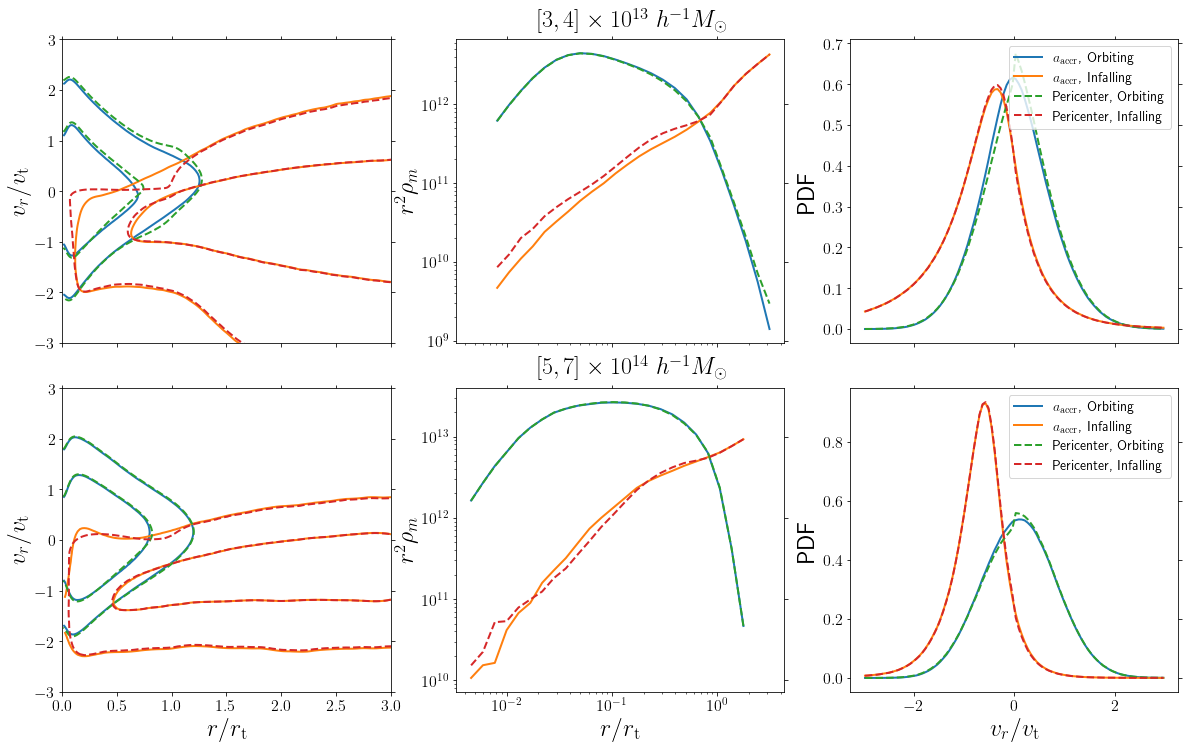}
    \caption{Comparison of the two classification schemes explored in this work for low (top row) and high (bottom row) mass haloes, as labelled. \textbf{Left:} 68 and 95\% contours of orbiting and infalling particles. The contours are consistent with only small deviations. \textbf{Center:} Orbiting and infalling density profiles. \textbf{Right:} One point functions of the radial velocities.}
    \label{fig:splits-comparison-2}
\end{figure*}

In the previous section, we established the orbiting--infall dichotomy of particles in the vicinity of dark matter haloes.  While our definition is physically motivated, it raises an obvious question: when does a particle transform from an infalling particle into an orbiting particle?  If we imagine a particle falling into a halo, an obvious choice for when this transition occurs is when the particle reaches its first pericentric passage, as suggested by \citet{diemer2022a}; but does such a definition in fact separate the two dynamically distinct components seen in Figures~\ref{fig:phase-space-2} and ~\ref{fig:Rafa-split}?  In this section, we test whether splitting particles by whether or not they have experienced their first pericentric passage is equivalent to splitting particles into the orbiting and infalling components identified in the previous section.  In principle, splitting particles by pericentric passage is simple: given the orbital radius $r(a)$, we need only determine whether the particle reached a minimum in its orbit.  In practice, we find that noise in the data, and particularly particles orbiting substructures, render this type of approach unfeasible.  Instead we turn to analyzing particle orbits in the $r$-$v_r$ plane. In this sense, our algorithm is simpler than that of \citet{diemer2022a}, which uses full 3D positions.

Figure \ref{fig:Edgar-split} shows an example of a typical particle orbit. The particle starts at a large distance from the halo with negative radial velocity, at some angle $\theta_0$ relative to the $r$-axis. It falls into the halo until it reaches its pericenter and keeps circling around in the $v_r$--$r$ space thereafter.  For orbiting particles, this circling happens about an average radius $r_c$ (`c' for ``center'') that can be roughly estimated as the average radius of the particle over the last dynamical time. That is, we define $r_c \equiv \avg{r}$, where $\avg{r}$ is calculated over the last dynamical time. If $\avg{r}>\rt$, we set $r_c = \rt$ instead. We find this restriction ensures that particles labelled orbiting are orbiting the central halo under consideration, rather than a nearby halo.  To distinguish orbiting from infalling particles, we calculate the angle swept from the $r$-axis as the particle traverses its orbit, using the point $(r_c,0)$ as our reference point.  The choice of reference point is critical: if this point falls outside the phase space loop corresponding to a particle's pericentric passage, the resulting angle need not be greater than $\pi$ (see e.g., Figure~\ref{fig:missclassified-orbit}).  In detail, the angle swept is computed as

\begin{equation}
    \Theta = \sum_i \theta_i,
\end{equation}

where $\theta_i$ is the signed angle swept between snapshot $i$ and snapshot $i+1$.  Clockwise angles are positive. The magnitude of the angle swept $\theta_i$ is given by $\theta_i = \arccos\left( \mathbf{\hat x}_i\cdot \mathbf{\hat x}_{i+1} \right)$. The snapshot indices start at $i=1$, but the indexed sum starts at $i=0$. We define $\mathbf{\hat x}_0 = \mathbf{\hat r}$, so that $\theta_0$ is the initial angle of the particle as shown in Figure~\ref{fig:Edgar-split}. A particle that is infalling will exhibit its first pericentric passage at $\Theta=\pi$.  

With this algorithm in hand, we define infalling particles as those with $\Theta < \pi$ and orbiting particles as those with $\Theta\geq \pi$. Figure \ref{fig:high-splits-comparison} shows the resulting distributions of orbiting (top row) and infalling (bottom row) particles in the $\aacc$--$\avg{v_r}$ plane for a high mass halo bin ($M\in [5,7]\times 10^{14}\ \hmsun$).  The corresponding figure for a low-mass halo bin is qualitatively similar (not shown). The color scale is logarithmic to highlight the tails of the distributions. The cyan line is the line used in section~\ref{sec:Rafa-split} to perform the orbiting/infall decomposition.  We see that the accretion-time based particle split is in good agreement with the pericentic-based particle split, though some differences become apparent at small radii.  When using the pericentric-based decomposition, both the orbiting and infall particle distributions exhibit a clear discontinuity at $\avg{v_r}=0$.  Focusing on the orbiting particles, the discontinuity at $\avg{v_r}=0$ corresponds to either a deficit of orbiting particles with negative radial velocities, or an over-abundance of orbiting particles with positive radial velocities (or both).  Looking at the infalling distribution, we see that the pericentric passage definition labels many particles with early accretion times as ``infalling''.  This suggests that our pericentric definition may incorrectly label some orbiting particles as infalling.  In Figure~\ref{fig:missclassified-orbit} we show a typical orbit for one of these particles, i.e., a particle labelled as infalling with an early accretion time.  Most such particles have qualitatively similar orbits.  Figure~\ref{fig:missclassified-orbit} makes it clear that the pericentric definition failed in this case: the radius $r_{\rm c}$ used to measure the swept angle $\Theta$ was slightly too large, and therefore the ``splashback'' portion of the angle swept got added up with the wrong sign.  Based on these results, we believe the orbiting and infall particle split described in section~\ref{sec:Rafa-split} is superior to this alternative pericentric-passage based definition.  We note that because the algorithm of \citet{diemer2022a} relies on 3D positions, it does not suffer from this specific failure mode. We postpone a detailed comparison between the \citet{diemer2022a} algorithm and that of section~\ref{sec:Rafa-split} to future work.

To quantify the level of agreement between our two classification schemes, we calculate the fraction of particles whose classification differs between the accretion-time method and the pericenteric-passage method, as a function of radius.  This fraction is less than 3\% (10\%) at all radii for our highest (lowest) halo mass bin. The largest deviations occur near the halo radius $\rt$.  Note that not only is this fraction small, as shown in Figure~\ref{fig:splits-comparison-2}, the phase space structure of the orbiting and infalling particles is very similar for the two classification schemes we employed. As usual, the top row corresponds to a low mass halo bin, while the bottom row corresponds to a high mass halo bin.   The left column shows the phase-space contours containing 68\% and 95\% of the orbiting and infall particles according to each of the two definitions, as labeled.  The 68\% contours are nearly identical between the two methods, with small differences becoming apparent in the 95\% contours.  Also shown are the particle densities projected along each of the two axis, i.e., density as a function of radius (center), and density as a function of radial velocity (right). As noted earlier, there is an obvious discontinuity at $v_r=0$ in the velocity distribution of orbiting particles when using the pericentric-passage definition. 

Based on these results, we arrive at two conclusions:
\begin{itemize}
    \item For the purposes of developing a theoretical model of large scale structure, the orbiting/infall split based on accretion times is superior to the pericentric-passage based definition above. This is because of the discontinuity at $v_r=0$ in the phase space distribution recovered using the latter definition. Such a discontinuity will necessarily complicate quantitative descriptions of the phase space structure of orbiting particles, impacting our ability to use halo-model descriptions of the velocity field of the Universe.
    \item While the accretion-based definition is superior to the pericentric-passage definition, the two are broadly consistent with each other.  In particular, while operationally it is better to rely on $\aacc$ and $\avg{v_r}$, we can still think of a particles as ``switching'' from infalling to orbiting when it reaches its first pericentric passage.
\end{itemize}

As emphasized in the introduction, these conclusions only apply with regards to the pericentric definition considered in this work.  A comparison to the pericentric passage definition of \citet{diemer2022a} will be presented in future work.	
	
\section{A Physically Sound Halo Definition and Its Implications}
\label{sec:implications-lss}

\subsection{A Physically Sound Halo Definition}
\label{sec:halo-def}

We have established that the dynamical structure and histories of particles in the vicinity of a dark matter halo exhibit strong bimodality.  Moreover, we have argued that these two components correspond to a population of particles orbiting around the halo, while the second population comprises particles that are infalling into the halo for the first time.  In light of this dichotomy, we propose to define haloes as follows: 

\it A dark matter halo is comprised of the collection of particles orbiting in their own self-generated potential. \rm  

We refer to haloes selected with this definition as \it physical haloes. \rm Contrary to traditional halo definitions, our proposed definition is firmly rooted in the dynamics and dynamical history of the particles in the halo.  Moreover, there is no part of our definition that is arbitrary.  For instance, if we were to adopt the pericentric-passage criterion for distinguishing orbiting from infalling particles, {\it the definition has in principle no free parameters}.\footnote{In practice, pericentric definitions still rely on an a priori choice of a radial scale for numerical efficiency and improving algorithmic performance.  However, this sensitivity is small.} For our preferred split, there is an ``arbitrary'' choice of radial threshold used to define $\aacc$.  However, our results are remarkably robust to the choice of threshold.  As an example, we changed the radial threshold used to define $\aacc$ by 50\% --- i.e., from $\rt$ to $0.5\rt$ or $1.5\rt$ --- and recomputed the orbiting profiles of our haloes.  The fraction of orbiting particles that are reclassified as infalling when applying these changes are 1.8\% and 1.4\% respectively, while the median absolute deviation between the resulting orbiting profiles is $\lesssim 0.5\%$ across all radii. It is for this reason that we refer to our selection as \it effectively parameter-free: \rm the choice of radius matters, but only at the percent level.

\subsection{Halo Percolation}

There remains two subtleties that need to be addressed when defining the new halo catalog.  The first concerns orbiting particles at large halo-centric radii.  If one does not take the necessary precautions, occasionally a single particle can be tagged as an orbiting particle of more than one halo, which doesn't make sense. To avoid this, we allow every particle to orbit \it at most \rm one halo by percolating the halo catalog.  We rank-order the haloes according to $\Mt$, and then select the orbiting particles belong to the most massive halo.  These particles are no longer allowed to orbit a less massive halo.  We then move on to the next most massive halo, and iterate the procedure until we go through the entire halo list.  This way, any one particle orbits at most one dark matter halo.

The second subtlety is that, in its current form, our newly constructed halo catalog still suffers from halo exclusion effects.  Specifically, our halo catalog is a subset of the original \rockstar\ halo catalog, which suffers from halo exclusion (i.e., haloes within a given distance of a more massive parent halo are automatically labelled subhaloes).  Our algorithm has led us to reject some $M_{200{\rm m}}$ haloes as unphysical (e.g., $M_{200{\rm m}}$ haloes orbiting a more massive structure outside the nominal halo boundary used by \rockstar), but we have not added any haloes back in.  In practice, we expect that there exists a population of infalling haloes in the vicinity of massive haloes which have not yet began orbiting around their more massive neighbor.  Consequently, these low mass haloes ought to be ``added back in'' as parent haloes.  In future work, we will determine the impact these haloes have, and characterize how the addition of these haloes, and in particular the removal of halo exclusion effects, impacts the halo--halo correlation function.

\subsection{The Orbiting Mass of a Halo}
\label{sec:orbiting-mass}

\begin{figure}
    \includegraphics[width=\linewidth]{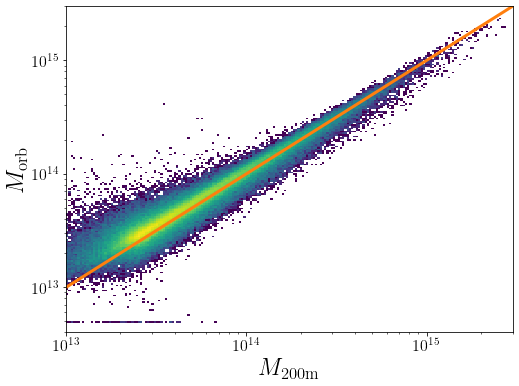}
    \caption{Relation between the orbiting mass and $M_{\rm 200m}$.  The objects with $\Morb<5\times10^{12}\ \hmsun$ are collapsed to that value for plotting purposes. The small orbiting mass of these objects suggests that they are not parent haloes since most of the particles originally assigned to them are orbiting other parent haloes.}
    \label{fig:morb-m200m}
\end{figure}

The orbiting mass of a halo $\Morb$ is the sum of the mass of all its orbiting particles. Unsurprisingly, it is tightly correlated with both $M_{200{\rm m}}$ and $\Mt$.  Figure~\ref{fig:morb-m200m} shows the orbiting mass of haloes as a function of $M_{200{\rm m}}$.  We model $P(\ln \Morb|M_{200{\rm m}})$ as a Gaussian distribution of mean
\begin{equation}
    \avg{\ln \Morb|M_{200{\rm m}}} = a\ln (M_{200{\rm m}}/M_p) + b
\end{equation}
and variance $\sigma^2$.  We choose $\Mp=5\times 10^{13}\ \hmsun$, which effectively decorrelates the parameters $a$, $b$, and $\sigma$. Fitting the data in Figure~\ref{fig:morb-m200m} for $M_{200{\rm m}}\geq 2\times 10^{13}\ \hMsun$ we find 
\begin{eqnarray}
a & = & 0.916 \pm 0.001 \\
b & = & 31.617 \pm 0.001 \\
\sigma & = & 0.163 \pm 0.001.
\end{eqnarray}
The corresponding relation between $\Morb$ and $\Mt$ is
\begin{eqnarray}
a' & = & 0.987 \pm 0.002 \\
b' & = & 31.633 \pm 0.001 \\
\sigma' & = & 0.215 \pm 0.001
\end{eqnarray}
where the pivot mass is now $\Mp=7\times10^{13}\ \hmsun$.\footnote{The pivot mass must change, since the numerical value of the mass of an object at the pivot mass changes depending on the halo definition.} The relationship between the orbiting mass and the mass within the splashback radius, $M_{\rm sp}$, is expected to be somewhat more complicated. It will certainly depend on the percentile of particle apocenters used to define $R_{\rm sp}$.  It is also not clear whether $M_{\rm sp}$ is larger or smaller than $M_{\rm orb}$: while $M_{\rm orb}$ should include all orbiting particles, which suggests $M_{\rm orb}$ should be larger than $M_{\rm sp}$, the splashback mass $M_{\rm sp}$ can have a non-negligible contribution from infalling particles, which might make $M_{\rm orb}$ smaller than $M_{\rm sp}$. We defer a detailed comparison of the orbiting and splashback masses to future work.

We find that our halo percolation results in a small number of massive $M_{200{\rm m}}$ haloes ($M_{200{\rm m}} \geq 10^{13}\ \hMsun)$ being assigned a very small orbiting mass ($\Morb \lesssim 5\times 10^{12}\ \hMsun$).  These objects are systems where a significant subset of the original halo particles have been identified as orbiting a more massive halo.  That is, these systems are in fact subhaloes of a larger system that were ``erroneously'' included as parent haloes in the original \rockstar\ halo catalog.\footnote{We put the word ``erroneously'' in quotations because the assignment of haloes as parent haloes or not is a definitional question.}

\subsection{The Mass Function of Physical haloes}

\begin{figure}
    \includegraphics[width=\linewidth]{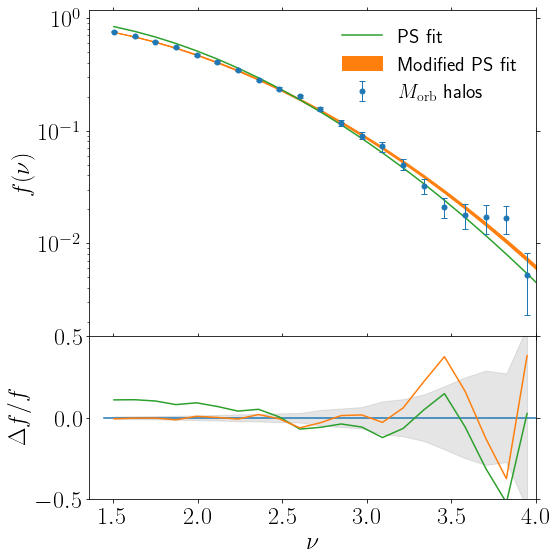}
    \caption{The halo mass function of $\Morb$ haloes is shown in blue points with error bars. The orange band shows the best fit Press--Schechter mass function. The bottom panel shows the ratio of best fit model to simulation data. Error bars are jackknife.}
    \label{fig:mass-function}
\end{figure}

In \cite{GarciaRozo2021} we showed that the halo mass function of $\Mt$ haloes is accurately described by the Press--Schechter formula \citep{Press-Schechter1974}. This mass function is of the form

\begin{equation}
    \frac{dn}{dm} = f(\sigma) \frac{\mean{\rho}_{\rm m}}{m} \frac{d \ln \sigma^{-1}}{dm}
\end{equation}

where $f(\sigma)$ is the multiplicity function and $\sigma(m)$ is the variance in the initial density fluctuation field. In the context of spherical collapse the multiplicity function is given by

\begin{equation}
    f(\sigma) = \sqrt{\frac{2}{\pi}} \frac{\dsc}{\sigma} \exp \left[ -\frac{\dsc^2}{2\sigma^2} \right]
\end{equation}

where $\deltasc$ is the overdensity threshold for collapse. The multiplicity function is also usually written in terms of the peak height $\nu \equiv \deltasc / \sigma(m)$. The value of the overdensity threshold predicted from linear theory at $z=0$ is $\deltasc=1.686$.  Importantly, in \citet{GarciaRozo2021} we fit for the critical threshold for collapse, finding the data favored the surprisingly low value $\deltasc=1.449$.  

Turning to the mass function of physical haloes, we find that a standard Press--Schechter fit results in $\deltasc=1.60$.  However, Figure~\ref{fig:mass-function} demonstrates that the Press--Schechter fit overestimates the number of low peak haloes by $5\%-10\%$.  To account for this difference, we assume the spherical threshold for collapse is mildly dependent on peak height.  Specifically, we assume that
\begin{equation}
    \deltasc(\nu_0) = \delta_{\rm sc, 0} + b \exp^{-\nu_0}.
\end{equation}
where $\nu_0 = \delta_{\rm sc, 0} / \sigma(m)$. We define the peak height 
\begin{equation}
    \nu \equiv \deltasc(\nu_0(m)) / \sigma(m).
    \label{eq:peak}
\end{equation}
The fraction of cells of volume $V=m/\bar \rho_{\rm m}$ in the initial density field that collapse into objects of mass greater than $m$ is given by
\begin{equation}
    F(m) = \int_{\deltasc}^{\infty} \sqrt{\frac{2}{\pi}} \frac{1}{\sigma} e^{-\frac{\delta^2}{2\sigma^2}} d\delta
\end{equation}
In terms of the redefined peak height $\nu$ (eq.~\ref{eq:peak}) we have
\begin{equation}
    F(M) = \int_{\nu}^{\infty} \sqrt{\frac{2}{\pi}} e^{-\frac{\nu'^2}{2}} d\nu'.
    \label{eq:F}
\end{equation}
Following the usual derivation from this point, we finally arrive at
\begin{equation}
    \frac{dn}{dm} dm = \sqrt{\frac{2}{\pi}} \frac{\mean{\rho}_m}{m^2} \nu \frac{d\ln{\nu}}{d\ln{m}} e^{-\frac{\nu^2}{2}} dm.
\end{equation}
That is, our final result is identical to the Press--Schechter expression when written in terms of our redefined peak height.

The resulting best fit model is shown in the orange band in Figure~\ref{fig:mass-function}, and is in excellent agreement with the data.  The best fit values of the parameters are $\delta_{\rm sc, 0}=1.55$ and $b=0.39$.  With these values the critical overdensity varies between 1.64 at $\nu=1.5$ (our lowest mass bin) to 1.56 at $\nu=4.0$ (our largest mass bin).

We emphasize that the range of spherical collapse thresholds we have recovered empirically ($1.64 \gtrsim \dsc \gtrsim 1.56$) is bounded by the range of spherical collapse thresholds predicted by the spherical collapse model.  Specifically, \citet{Shapiroetal99} pointed out that the traditional spherical collapse threshold of $\dsc=1.68$ is due to the assumption that virialization occurs when an infalling shell crosses zero.  However, if one assumes instead that spherical collapse is halted at the time when an infalling shell crosses the halo boundary, $\dsc$ is lowered to $\dsc=1.52$.  In practice, these two extremes limit the effective threshold for spherical collapse, so the fact that our results are bounded by these two limiting cases is remarkable.

These results demonstrate that the mass function of physical haloes can be more easily described than that derived using more standard halo definitions.  Specifically, the inclusion of a mild mass dependence on the definition of peak height is a minor modification of the standard Press--Schechter prediction, particularly when compared to common parametric fits proposed in the literature for traditionally defined haloes \citep[e.g.,][]{jenkins_01, shethtormen, tinker_08,watsonetal13}. Thus, our results suggest that the degrees of freedom introduced in these models are in fact accounting for unnecessary complications that arise at least in part because of the choice of halo definition.  In future work we will determine whether our proposed halo definition also renders the resulting halo mass function more Universal across redshift and cosmology \citep[e.g.,][]{reed_03, tinker_08, more_11_fof, despali_16, Diemer20}.  If so, this would considerably strengthen the argument in favor of our newly proposed halo definition.

\subsection{Halo Bias and the Peak-Background Split}

\begin{figure}
    \includegraphics[width=\linewidth]{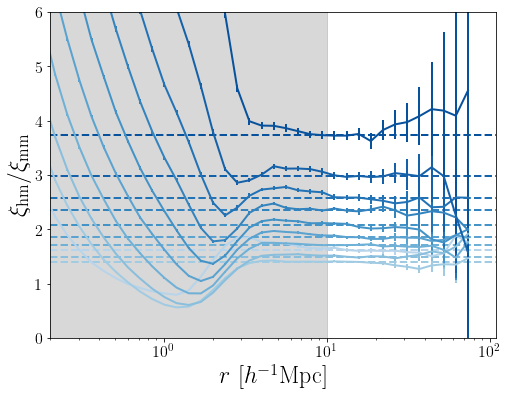}
    \caption{The solid lines show the ratio of halo--matter correlation function to matter--matter correlation function for multiple mass bins.  The dashed horizontal lines show the large-scale clustering amplitude measured from the ratios. We use scales larger than $10\ \hMpc$ to compute the clustering bias. The error bars are obtained by jackknifing the simulation volume.}
    \label{fig:xihm-ximm-ratio}
\end{figure}

We measure the clustering bias of physical haloes by computing the ratio $\xihm/\ximm$, where $\xihm$ is the halo--mass correlation function of physical haloes in a given mass bin, and $\ximm$ is the mass correlation function measured in the simulation.  Figure~\ref{fig:xihm-ximm-ratio} demonstrates that this ratio is constant at large scales.  Consequently, we estimate the clustering bias as the inverse-variance weighted mean $\xihm/\ximm$ ratio on scales $r\geq 10\ \hMpc$.

We compare our measurements to the predictions from the peak--background split model.  The peak-background split formalism predicts that if the halo mass function follows the Press--Schechter form, then the halo bias as a function of peak height takes the form
\begin{equation}
    b_{\rm PB} (\nu) = 1 + \frac{\nu^2 - 1}{\deltasc}.
\end{equation}
We find that our redefinition of halo peak height using a mass-dependent collapse threshold does not impact this prediction \citep[see also][]{desjacquesetal18}.  

\begin{figure}
    \includegraphics[width=\linewidth]{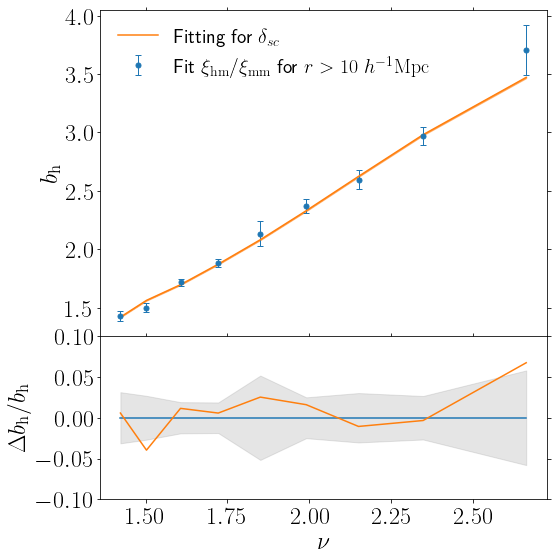}
    \caption{The top panel shows the halo bias measured from the simulation in blue point with error bars. The orange band is the \it prediction \rm from the peak-background split halo bias based on the mass function data and \it is not a fit. \rm  The bottom panel shows the percent different of our model prediction to the data. Error bars are jackknife.}
    \label{fig:halo-bias}
\end{figure}

Figure \ref{fig:halo-bias} shows a comparison between the halo bias measured in simulations and the peak-background split prediction based on our modified halo peak height.  The predicted bias is obtained by calculating the peak height of each individual halo based on its mass.  Using the peak--background split, we assign a clustering bias to each halo, and then average over all haloes in a mass bin to arrive at our final prediction.  Remarkably, the peak-background split is statistically consistent with the data ($\chi^2/dof=26.88/19$), despite the statistical uncertainties in the measurements falling consistently below 5\%.  This is to be compared to the $\approx 10\%-20\%$ accuracy achieved using traditional halo definitions, including $\Mt$ \citep{maneraetal10,Hoffmann2015,GarciaRozo2021}. 

\subsection{Is the Proposed Definition Observable?}
\label{sec:observable}

We have argued that our proposed halo definition is physically motivated, is effectively parameter free, and that the resulting halo statistics are well described by long-standing physical models. One might worry all these advantages would be for naught if the resulting mass definition is not observable. However, this is \it not \rm the case. Or, at the very least, not any more so than any other standard halo definition.

First, there is the obvious fact that because dark matter is dark, halo mass is never directly observed.  Rather, it must be inferred from other observational signatures.  This requires that one develop accurate model predictions for how the specific observables in question relate to the underlying halo mass.  For example, given a sample of clusters or other such halo proxy (e.g., central galaxies), the lensing signal of these object is predicted as a line-of-sight integral of the halo--mass correlation function.  Consequently, given a model for the halo--mass correlation function --- which we will introduce in future work --- our proposed halo mass is just as easily measured using gravitational lensing as any other standard halo definition. 

Turning to galaxy dynamics, a recent paper by \citet{aungetal22} demonstrated that: 1) the basic orbiting--infall decomposition can also be applied to simulated galaxies; and 2) the line-of-sight velocity structure of each of these two types of galaxies can be accurately described using simple fitting functions.  While \citet{aungetal22} based their work on $M_{200{\rm m}}$ haloes, one can update their model so that the fundamental scaling relation ties the velocity dispersion of galaxies not to $M_{200{\rm m}}$, but rather to $\Morb$.  In fact, given the physical model in \citet{aungetal22}, one might expect this scaling relation to be tighter when using $\Morb$.

In summary, we do not see any obstacle for inferring $\Morb$ from observational signatures that is not already present for standard halo mass definitions.

\section{Summary and Discussion}
\label{sec:summary}

This paper proposes a new definition of dark matter haloes.  Specifically:

\it A dark matter halo is comprised of the collection of particles orbiting their own self-generated potential. \rm  

This definition is based on the observation that the phase space distribution of dark matter particles in the vicinity of haloes is obviously bimodal (Figures~\ref{fig:phase-space-1} and \ref{fig:phase-space-2}).  We argued the two components can be interpreted as: 1) particles orbiting the dark matter halo;  and 2) particles falling into the halo for the first time. Because these two particle distributions overlap in phase space, there is no spherical overdensity definition that can adequately discriminate between the two populations.  Consequently, \it all current halo definitions necessarily fail at cleanly separating orbiting from infalling particles. \rm  

We then demonstrated that we can distinguish orbiting from infalling particles using particle dynamical histories (Figure~\ref{fig:Rafa-split}).  While the definition requires using a specific radial scale for the purposes of defining a particle's accretion time, a particle's classification is  robust to changes in this scale (see Section~\ref{sec:halo-def}).  Consequently, our classification is effectively parameter free.  We have also verified that orbiting and infalling particles can be differentiated based on whether a particle has experienced its first halo pericentric passage (Section~\ref{sec:Edgar-split}), a definition that is in principle truly parameter-free.  However, our implementation of this alternative definition was nosier than our fiducial definition, and resulted in undesirable discontinuities in the phase space distribution of orbiting particles.  Consequently, we prefer our fiducial accretion-time based definition.  These conclusions are specific to our particular implementation of the pericentric passage definition, and they do not extend to the selection algorithm developed by \citet{diemer2022a}.

Having demonstrated that the particles in the vicinity of a halo can be robustly identified as either orbiting or infalling, we proceeded to define haloes as the collection of all particles orbiting around their self-generated potential. We argue in section~\ref{sec:observable} that this new definition is equivalent to standard halo definitions in terms of how easy it is (or not) to estimate a halo's mass from observational data. 

We then characterized the basic statistical properties of the halo catalogs produced by our halo definition. In particular, we demonstrated that the halo mass function can be accurately fit using the standard Press--Schechter formula (Figure~\ref{fig:mass-function}), but doing so requires a threshold barrier $\dsc$ that decreases slowly with mass. Specifically, we find $\dsc$ decreases from $\dsc=1.64$ at $\nu\approx 1.5$ to $\dsc=1.56$ at $\nu\approx 4$.  These numbers fall within the limiting values of  $\dsc=1.52$ and $\dsc=1.68$, calculated using spherical collapse, and assume that shell virialization occurs either when a shell first enters the halo ($\dsc=1.52$) or when a shell crosses $r=0$ \citep[$\dsc=1.68$,][]{Shapiroetal99}.

We also calculated the cluster bias of our haloes as the ratio $b\equiv \xihm/\ximm$ on scales $r\geq 10\ \hMpc$ (Figure~\ref{fig:xihm-ximm-ratio}).  Figure~\ref{fig:halo-bias} compares our recovered halo bias measurements to predictions from the peak--background split model.  Despite the comparatively large error bars, the agreement is remarkable, being no worse than $\approx 5\%$ across the entire mass range.  This is in stark contrast with the behavior seen with standard halo definitions, for which the peak background split is accurate only at the $10\%$--$20\%$ level, depending on the mass.


Figures~\ref{fig:phase-space-1}, \ref{fig:phase-space-2}, and \ref{fig:Rafa-split} provide a compelling argument as to why our proposed definition should be favored, irrespective of any theoretical benefits. Unlike traditional halo definitions, our proposal is firmly rooted in the dynamics of halo particles and is effectively parameter free.  The primary down side of our proposed definition is that it requires the use of the full particle history for every particle in the simulation.  On the flip side, the resulting halo statistics are remarkably simple to describe: the mass function of physical haloes is Press--Schechter, and their clustering bias is accurately predicted by the peak--background split model.  Moreover, the recovered collapse thresholds make sense.

In a future paper we will characterize the halo--mass correlation function of halos selected using our proposed definition.  We will demonstrate that the resulting halo model is self-consistent, simple, physically motivated, and accurate.  In short, those results provide further evidence that our halo mass definition is demonstrably better suited for describing the large scale structure of the Universe.  To make practical use of these advantages, however, will require developing fast algorithms that can accurately approximate the computationally cumbersome definition developed here.  With a fast algorithm in hand, we will be able to characterize the resulting halo statistics as a function of cosmology and redshift in large simulation suites.

\section*{Acknowledgement}

We would like to acknowledge Chunhao To and Andres Salcedo for comments that improved both the clarity and substance of this manuscript. RG, ES, and ER are supported by DOE grants DE-SC0009913. ER is also supported by NSF grant 2009401.

\bibliographystyle{mnras}
\bibliography{database.bib}

\appendix
\section{Convergence tests}
\label{app:convergence}

\begin{figure*}
    \includegraphics[width=0.49\textwidth]{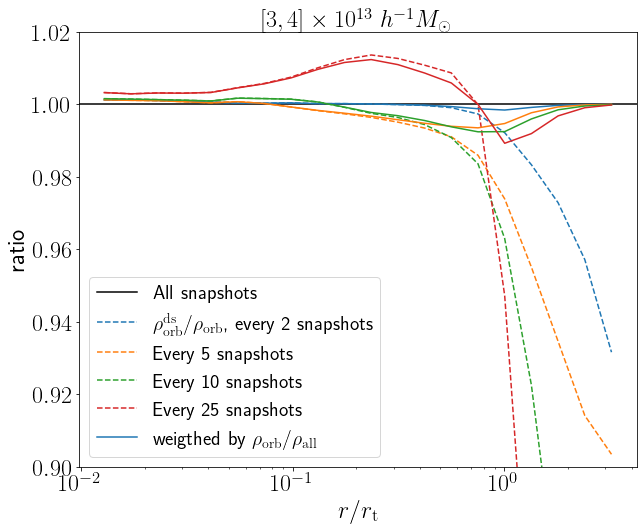}
    \includegraphics[width=0.49\textwidth]{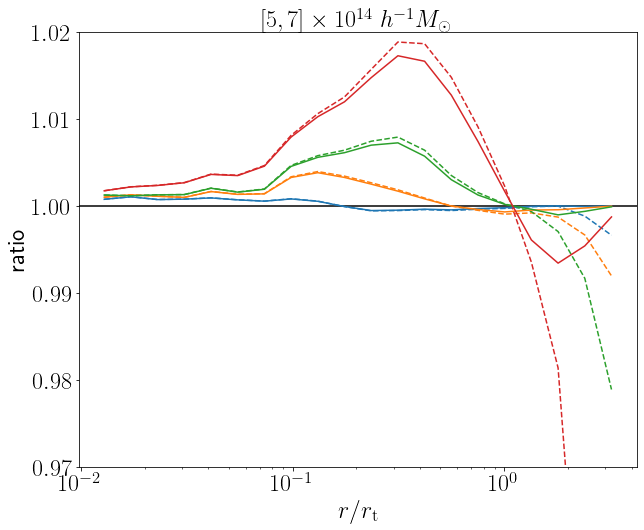}
    \caption{Dashed lines show the ratios of orbiting density profiles obtained by performing the split with varying number of snapshots to the orbiting profile calculated with all 100 available snapshots. The solid lines are calculated using equation~\ref{eq:ratio}, and demonstrate that uncertainties associated with time sampling are at the $\approx 1\%$ level relative to the full matter density profile. Small (high) mass bin shown on  the left (right).}
    \label{fig:convergence}
\end{figure*}

We test the numerical convergence of our fiducial (i.e., accretion-based) algorithm by performing the orbiting/infall split using a varying number of snapshots. We start by using the full set of 100 snapshots.  This is our fiducial setup, and what is used for all of the results presented in this paper.  We then we generate multiple sets of snapshots by downsampling the full set by skipping $i$ snapshots for values of $i \in \{2, 5, 10, 25\}$. For each of these sets we run our classification algorithm and compute the corresponding orbiting profile.

Figure~\ref{fig:convergence} shows the impact of time resolution on the accuracy of the orbiting profile.  The dashed lines show the ratio of the orbiting profile calculated with a downsampled set of snapshots $\rho_{\rm orb}^{\rm ds}$ to the profile calculated with the full set $\rho_{\rm orb}$, as labelled.  The left panel corresponds to our lowest mass bin, and the right panel to our highest $\Morb$ bin.  We see there is excellent agreement in the orbiting profiles for the high mass haloes across the spatial scales of interest. In the case of the low mass haloes, small ($\lesssim 10\%$) differences in the orbiting profiles become apparent at large radii ($r\gtrsim \rt$).  However, at these scales, the orbiting component is strongly subdominant to the infall component.  To demonstrate this, the solid lines in Figure~\ref{fig:convergence} plot the quantity 
\begin{equation}
    r = 1 + \frac{ \rho^{\rm ds}_{\rm orb} - \rho_{\rm orb} }{\rho_{\rm tot}}.
    \label{eq:ratio}
\end{equation}
That is, the deviation from zero between the original and downsampled profiles is expressed in units of the \it total \rm density.  Figure~\ref{fig:convergence} shows the uncertainties associated with time sampling correspond to $\approx 1\%$ of the total matter profile.  It is interesting to note that our accretion-time based algorithm can robustly distinguish orbiting from infalling particles with only a few snapshots.

\begin{figure*}
    \includegraphics[width=\textwidth]{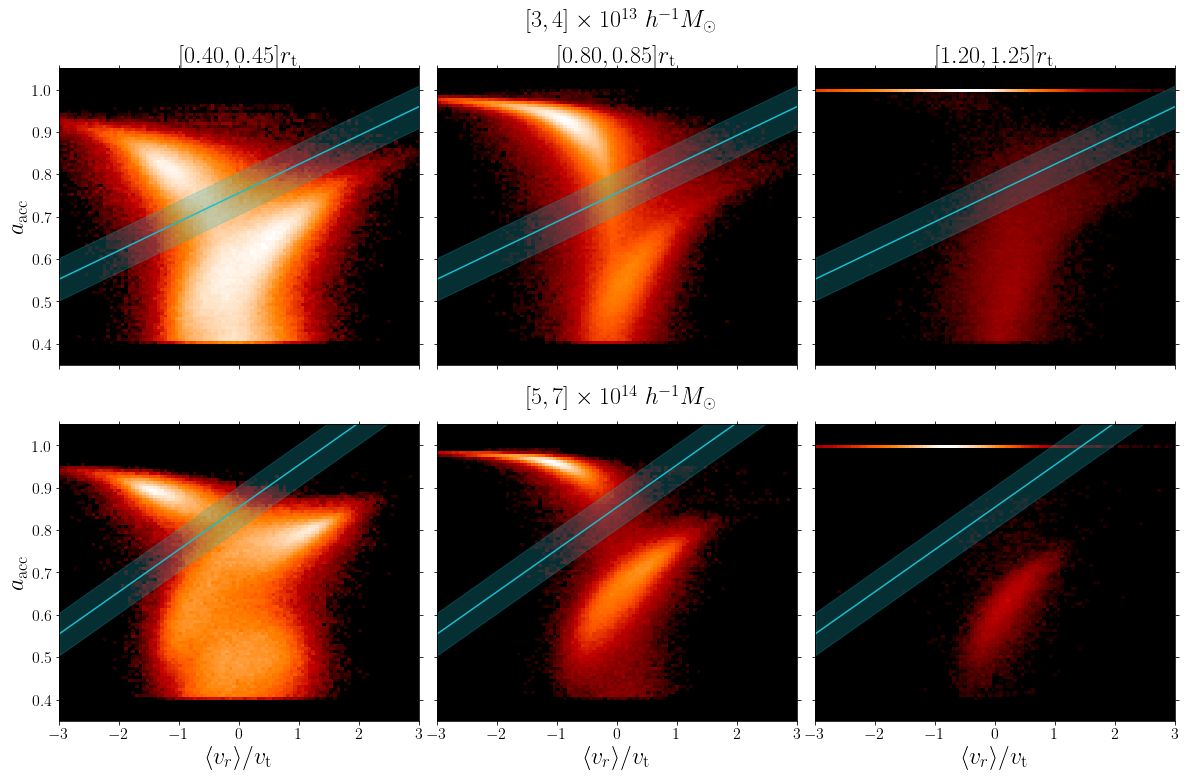}
    \caption{Same figure as Fig. \ref{fig:Rafa-split} but in logarithmic color scale. The logarithmic scale reveals that the orbiting particles go beyond the halo radius $\rt$ and that the orbiting and infalling populations have a small overlap not easily visible in linear scale.}
    \label{fig:Rafa-split-log}
\end{figure*}

To end, Figure~\ref{fig:Rafa-split-log} reproduces Figure~\ref{fig:Rafa-split}, only the particle density is now plotted using a logarithmic scale.  This figure highlights the remaining overlap between the orbiting and infall components present in the $\aacc$--$\avg{v_r}$ space.

\label{lastpage}

\end{document}

%% file: authorlist.tex
\author[Garcia et al.]{Rafael Garc\'ia$^1$\thanks{E-mail: rgarciamar@email.arizona.edu}, Edgar Salazar$^1$, Eduardo Rozo$^{1}$, Susmita Adhikari$^2$, Han Aung$^3$, \and Benedikt Diemer$^4$, Daisuke Nagai$^5$, Brandon Wolfe$^1$.
\\
$^{1}$Department of Physics, University of Arizona, Tucson, AZ 85721, USA \\
$^{2}$Department of Physics, Indian Institute of Science Education and Research, Homi Bhaba Road, Pashan, Pune 411008, India\\
$^{3}$Centre for Astrophysics and Planetary Science, Racah Institute of Physics, The Hebrew University, Jerusalem 91904, Israel\\
$^4$Department of Astronomy, University of Maryland, College Park, MD 20742, USA\\
$^{5}$Department of Physics, Yale University, New Haven, CT 06520, USA
}

%% file: main.bbl
\begin{thebibliography}{}
\makeatletter
\relax
\def\mn@urlcharsother{\let\do\@makeother \do\$\do\&\do\#\do\^\do\_\do\%\do\~}
\def\mn@doi{\begingroup\mn@urlcharsother \@ifnextchar [ {\mn@doi@}
  {\mn@doi@[]}}
\def\mn@doi@[#1]#2{\def\@tempa{#1}\ifx\@tempa\@empty \href
  {http://dx.doi.org/#2} {doi:#2}\else \href {http://dx.doi.org/#2} {#1}\fi
  \endgroup}
\def\mn@eprint#1#2{\mn@eprint@#1:#2::\@nil}
\def\mn@eprint@arXiv#1{\href {http://arxiv.org/abs/#1} {{\tt arXiv:#1}}}
\def\mn@eprint@dblp#1{\href {http://dblp.uni-trier.de/rec/bibtex/#1.xml}
  {dblp:#1}}
\def\mn@eprint@#1:#2:#3:#4\@nil{\def\@tempa {#1}\def\@tempb {#2}\def\@tempc
  {#3}\ifx \@tempc \@empty \let \@tempc \@tempb \let \@tempb \@tempa \fi \ifx
  \@tempb \@empty \def\@tempb {arXiv}\fi \@ifundefined
  {mn@eprint@\@tempb}{\@tempb:\@tempc}{\expandafter \expandafter \csname
  mn@eprint@\@tempb\endcsname \expandafter{\@tempc}}}

\bibitem[\protect\citeauthoryear{{Adhikari}, {Dalal}  \&
  {Chamberlain}}{{Adhikari} et~al.}{2014}]{adhikari2014}
{Adhikari} S.,  {Dalal} N.,   {Chamberlain} R.~T.,  2014, \mn@doi [Journal of
  Cosmology and Astro-Particle Physics] {10.1088/1475-7516/2014/11/019}, \href
  {https://ui.adsabs.harvard.edu/\#abs/2014JCAP...11..019A} {2014, 019}

\bibitem[\protect\citeauthoryear{{Adhikari}, {Dalal}, {More}  \&
  {Wetzel}}{{Adhikari} et~al.}{2019}]{adhikari_19}
{Adhikari} S.,  {Dalal} N.,  {More} S.,   {Wetzel} A.,  2019, \mn@doi [\apj]
  {10.3847/1538-4357/ab1a39}, \href
  {https://ui.adsabs.harvard.edu/abs/2019ApJ...878....9A} {878, 9}

\bibitem[\protect\citeauthoryear{{Adhikari} et~al.,}{{Adhikari}
  et~al.}{2021}]{adhikari_21}
{Adhikari} S.,  et~al., 2021, \mn@doi [\apj] {10.3847/1538-4357/ac0bbc}, \href
  {https://ui.adsabs.harvard.edu/abs/2021ApJ...923...37A} {923, 37}

\bibitem[\protect\citeauthoryear{{Aung}, {Nagai}, {Rozo}  \&
  {Garc{\'\i}a}}{{Aung} et~al.}{2021}]{Aung2021}
{Aung} H.,  {Nagai} D.,  {Rozo} E.,   {Garc{\'\i}a} R.,  2021, \mn@doi [\mnras]
  {10.1093/mnras/staa3994}, \href
  {https://ui.adsabs.harvard.edu/abs/2021MNRAS.502.1041A} {502, 1041}

\bibitem[\protect\citeauthoryear{{Aung}, {Nagai}, {Rozo}, {Wolfe}  \&
  {Adhikari}}{{Aung} et~al.}{2022}]{aungetal22}
{Aung} H.,  {Nagai} D.,  {Rozo} E.,  {Wolfe} B.,   {Adhikari} S.,  2022, arXiv
  e-prints, \href {https://ui.adsabs.harvard.edu/abs/2022arXiv220413131A} {p.
  arXiv:2204.13131}

\bibitem[\protect\citeauthoryear{{Bah{\'e}}, {McCarthy}, {Balogh}  \&
  {Font}}{{Bah{\'e}} et~al.}{2013}]{bahe_13}
{Bah{\'e}} Y.~M.,  {McCarthy} I.~G.,  {Balogh} M.~L.,   {Font} A.~S.,  2013,
  \mn@doi [\mnras] {10.1093/mnras/stt109}, \href
  {https://ui.adsabs.harvard.edu/abs/2013MNRAS.430.3017B} {430, 3017}

\bibitem[\protect\citeauthoryear{{Bakels}, {Ludlow}  \& {Power}}{{Bakels}
  et~al.}{2021}]{bakels_21}
{Bakels} L.,  {Ludlow} A.~D.,   {Power} C.,  2021, \mn@doi [\mnras]
  {10.1093/mnras/staa3979}, \href
  {https://ui.adsabs.harvard.edu/abs/2021MNRAS.501.5948B} {501, 5948}

\bibitem[\protect\citeauthoryear{{Balogh}, {Navarro}  \& {Morris}}{{Balogh}
  et~al.}{2000}]{Baloghetal00}
{Balogh} M.~L.,  {Navarro} J.~F.,   {Morris} S.~L.,  2000, \mn@doi [\apj]
  {10.1086/309323}, \href
  {https://ui.adsabs.harvard.edu/abs/2000ApJ...540..113B} {540, 113}

\bibitem[\protect\citeauthoryear{{Banerjee}, {Adhikari}, {Dalal}, {More}  \&
  {Kravtsov}}{{Banerjee} et~al.}{2020}]{banerjeeetal20}
{Banerjee} A.,  {Adhikari} S.,  {Dalal} N.,  {More} S.,   {Kravtsov} A.,  2020,
  \mn@doi [\jcap] {10.1088/1475-7516/2020/02/024}, \href
  {https://ui.adsabs.harvard.edu/abs/2020JCAP...02..024B} {2020, 024}

\bibitem[\protect\citeauthoryear{{Behroozi}, {Wechsler}  \& {Wu}}{{Behroozi}
  et~al.}{2013}]{Behroozi2013}
{Behroozi} P.~S.,  {Wechsler} R.~H.,   {Wu} H.-Y.,  2013, \mn@doi [\apj]
  {10.1088/0004-637X/762/2/109}, \href
  {http://adsabs.harvard.edu/abs/2013ApJ...762..109B} {762, 109}

\bibitem[\protect\citeauthoryear{{Behroozi}, {Wechsler}, {Lu}, {Hahn}, {Busha},
  {Klypin}  \& {Primack}}{{Behroozi} et~al.}{2014}]{behroozi_14}
{Behroozi} P.~S.,  {Wechsler} R.~H.,  {Lu} Y.,  {Hahn} O.,  {Busha} M.~T.,
  {Klypin} A.,   {Primack} J.~R.,  2014, \mn@doi [\apj]
  {10.1088/0004-637X/787/2/156}, \href
  {http://adsabs.harvard.edu/abs/2014ApJ...787..156B} {787, 156}

\bibitem[\protect\citeauthoryear{{Cooray} \& {Sheth}}{{Cooray} \&
  {Sheth}}{2002}]{Cooray-Sheth}
{Cooray} A.,  {Sheth} R.,  2002, \mn@doi [\physrep]
  {10.1016/S0370-1573(02)00276-4}, \href
  {http://adsabs.harvard.edu/abs/2002PhR...372....1C} {372, 1}

\bibitem[\protect\citeauthoryear{{Cuesta}, {Prada}, {Klypin}  \&
  {Moles}}{{Cuesta} et~al.}{2008}]{cuesta_08}
{Cuesta} A.~J.,  {Prada} F.,  {Klypin} A.,   {Moles} M.,  2008, \mn@doi
  [\mnras] {10.1111/j.1365-2966.2008.13590.x}, \href
  {http://adsabs.harvard.edu/abs/2008MNRAS.389..385C} {389, 385}

\bibitem[\protect\citeauthoryear{{Desjacques}, {Jeong}  \&
  {Schmidt}}{{Desjacques} et~al.}{2018}]{desjacquesetal18}
{Desjacques} V.,  {Jeong} D.,   {Schmidt} F.,  2018, \mn@doi [\physrep]
  {10.1016/j.physrep.2017.12.002}, \href
  {https://ui.adsabs.harvard.edu/abs/2018PhR...733....1D} {733, 1}

\bibitem[\protect\citeauthoryear{{Despali}, {Giocoli}, {Angulo}, {Tormen},
  {Sheth}, {Baso}  \& {Moscardini}}{{Despali} et~al.}{2016}]{despali_16}
{Despali} G.,  {Giocoli} C.,  {Angulo} R.~E.,  {Tormen} G.,  {Sheth} R.~K.,
  {Baso} G.,   {Moscardini} L.,  2016, \mn@doi [\mnras]
  {10.1093/mnras/stv2842}, \href
  {http://adsabs.harvard.edu/abs/2016MNRAS.456.2486D} {456, 2486}

\bibitem[\protect\citeauthoryear{{Diemand} \& {Kuhlen}}{{Diemand} \&
  {Kuhlen}}{2008}]{diemand_08}
{Diemand} J.,  {Kuhlen} M.,  2008, \mn@doi [\apjl] {10.1086/589688}, \href
  {http://adsabs.harvard.edu/abs/2008ApJ...680L..25D} {680, L25}

\bibitem[\protect\citeauthoryear{{Diemer}}{{Diemer}}{2017}]{diemer2017}
{Diemer} B.,  2017, \mn@doi [\apjs] {10.3847/1538-4365/aa799c}, \href
  {http://adsabs.harvard.edu/abs/2017ApJS..231....5D} {231, 5}

\bibitem[\protect\citeauthoryear{{Diemer}}{{Diemer}}{2020}]{Diemer20}
{Diemer} B.,  2020, \mn@doi [\apj] {10.3847/1538-4357/abbf52}, \href
  {https://ui.adsabs.harvard.edu/abs/2020ApJ...903...87D} {903, 87}

\bibitem[\protect\citeauthoryear{{Diemer}}{{Diemer}}{2021}]{Diemer2021}
{Diemer} B.,  2021, \mn@doi [\apj] {10.3847/1538-4357/abd947}, \href
  {https://ui.adsabs.harvard.edu/abs/2021ApJ...909..112D} {909, 112}

\bibitem[\protect\citeauthoryear{{Diemer}}{{Diemer}}{2022a}]{diemer2022b}
{Diemer} B.,  2022a, arXiv e-prints, \href
  {https://ui.adsabs.harvard.edu/abs/2022arXiv220503420D} {p. arXiv:2205.03420}

\bibitem[\protect\citeauthoryear{{Diemer}}{{Diemer}}{2022b}]{diemer2022a}
{Diemer} B.,  2022b, \mn@doi [\mnras] {10.1093/mnras/stac878}, \href
  {https://ui.adsabs.harvard.edu/abs/2022MNRAS.513..573D} {513, 573}

\bibitem[\protect\citeauthoryear{{Diemer} \& {Kravtsov}}{{Diemer} \&
  {Kravtsov}}{2014}]{diemerkravtsov14}
{Diemer} B.,  {Kravtsov} A.~V.,  2014, \mn@doi [\apj]
  {10.1088/0004-637X/789/1/1}, \href
  {http://adsabs.harvard.edu/abs/2014ApJ...789....1D} {789, 1}

\bibitem[\protect\citeauthoryear{{Diemer}, {More}  \& {Kravtsov}}{{Diemer}
  et~al.}{2013}]{diemer2013a}
{Diemer} B.,  {More} S.,   {Kravtsov} A.~V.,  2013, \mn@doi [\apj]
  {10.1088/0004-637X/766/1/25}, \href
  {https://ui.adsabs.harvard.edu/abs/2013ApJ...766...25D} {766, 25}

\bibitem[\protect\citeauthoryear{{Fillmore} \& {Goldreich}}{{Fillmore} \&
  {Goldreich}}{1984}]{fillmore_84}
{Fillmore} J.~A.,  {Goldreich} P.,  1984, \mn@doi [\apj] {10.1086/162070},
  \href {http://adsabs.harvard.edu/abs/1984ApJ...281....1F} {281, 1}

\bibitem[\protect\citeauthoryear{{Fong} \& {Han}}{{Fong} \&
  {Han}}{2021}]{fong_21}
{Fong} M.,  {Han} J.,  2021, \mn@doi [\mnras] {10.1093/mnras/stab259}, \href
  {https://ui.adsabs.harvard.edu/abs/2021MNRAS.503.4250F} {503, 4250}

\bibitem[\protect\citeauthoryear{{Fukushige} \& {Makino}}{{Fukushige} \&
  {Makino}}{2001}]{Fukushige01}
{Fukushige} T.,  {Makino} J.,  2001, \mn@doi [\apj] {10.1086/321666}, \href
  {https://ui.adsabs.harvard.edu/abs/2001ApJ...557..533F} {557, 533}

\bibitem[\protect\citeauthoryear{{Garc{\'\i}a} \& {Rozo}}{{Garc{\'\i}a} \&
  {Rozo}}{2019}]{GarciaRozo2019}
{Garc{\'\i}a} R.,  {Rozo} E.,  2019, \mn@doi [\mnras] {10.1093/mnras/stz2458},
  \href {https://ui.adsabs.harvard.edu/abs/2019MNRAS.489.4170G} {489, 4170}

\bibitem[\protect\citeauthoryear{{Garc{\'\i}a}, {Rozo}, {Becker}  \&
  {More}}{{Garc{\'\i}a} et~al.}{2021}]{GarciaRozo2021}
{Garc{\'\i}a} R.,  {Rozo} E.,  {Becker} M.~R.,   {More} S.,  2021, \mn@doi
  [\mnras] {10.1093/mnras/stab1317}, \href
  {https://ui.adsabs.harvard.edu/abs/2021MNRAS.505.1195G} {505, 1195}

\bibitem[\protect\citeauthoryear{{Gill}, {Knebe}  \& {Gibson}}{{Gill}
  et~al.}{2005}]{gill_05}
{Gill} S.~P.~D.,  {Knebe} A.,   {Gibson} B.~K.,  2005, \mn@doi [\mnras]
  {10.1111/j.1365-2966.2004.08562.x}, \href
  {http://adsabs.harvard.edu/abs/2005MNRAS.356.1327G} {356, 1327}

\bibitem[\protect\citeauthoryear{{Gunn} \& {Gott}}{{Gunn} \&
  {Gott}}{1972}]{gunn_72}
{Gunn} J.~E.,  {Gott} III J.~R.,  1972, \mn@doi [\apj] {10.1086/151605}, \href
  {http://adsabs.harvard.edu/abs/1972ApJ...176....1G} {176, 1}

\bibitem[\protect\citeauthoryear{{Hamabata}, {Oguri}  \&
  {Nishimichi}}{{Hamabata} et~al.}{2019}]{Hamabata2019}
{Hamabata} A.,  {Oguri} M.,   {Nishimichi} T.,  2019, \mn@doi [\mnras]
  {10.1093/mnras/stz2227}, \href
  {https://ui.adsabs.harvard.edu/abs/2019MNRAS.489.1344H} {489, 1344}

\bibitem[\protect\citeauthoryear{{Hoffmann}, {Bel}  \&
  {Gazta{\~n}aga}}{{Hoffmann} et~al.}{2015}]{Hoffmann2015}
{Hoffmann} K.,  {Bel} J.,   {Gazta{\~n}aga} E.,  2015, \mn@doi [\mnras]
  {10.1093/mnras/stv702}, \href
  {http://adsabs.harvard.edu/abs/2015MNRAS.450.1674H} {450, 1674}

\bibitem[\protect\citeauthoryear{{Jenkins}, {Frenk}, {White}, {Colberg},
  {Cole}, {Evrard}, {Couchman}  \& {Yoshida}}{{Jenkins}
  et~al.}{2001}]{jenkins_01}
{Jenkins} A.,  {Frenk} C.~S.,  {White} S.~D.~M.,  {Colberg} J.~M.,  {Cole} S.,
  {Evrard} A.~E.,  {Couchman} H.~M.~P.,   {Yoshida} N.,  2001, \mn@doi [\mnras]
  {10.1046/j.1365-8711.2001.04029.x}, \href
  {http://adsabs.harvard.edu/abs/2001MNRAS.321..372J} {321, 372}

\bibitem[\protect\citeauthoryear{{Lacey} \& {Cole}}{{Lacey} \&
  {Cole}}{1994}]{lacey_94}
{Lacey} C.,  {Cole} S.,  1994, \mn@doi [\mnras] {10.1093/mnras/271.3.676},
  \href {https://ui.adsabs.harvard.edu/abs/1994MNRAS.271..676L} {271, 676}

\bibitem[\protect\citeauthoryear{{Mamon}, {Sanchis}, {Salvador-Sol{\'e}}  \&
  {Solanes}}{{Mamon} et~al.}{2004}]{Mamonetal04}
{Mamon} G.~A.,  {Sanchis} T.,  {Salvador-Sol{\'e}} E.,   {Solanes} J.~M.,
  2004, \mn@doi [\aap] {10.1051/0004-6361:20034155}, \href
  {https://ui.adsabs.harvard.edu/abs/2004A&A...414..445M} {414, 445}

\bibitem[\protect\citeauthoryear{{Manera}, {Sheth}  \& {Scoccimarro}}{{Manera}
  et~al.}{2010}]{maneraetal10}
{Manera} M.,  {Sheth} R.~K.,   {Scoccimarro} R.,  2010, \mn@doi [\mnras]
  {10.1111/j.1365-2966.2009.15921.x}, \href
  {https://ui.adsabs.harvard.edu/abs/2010MNRAS.402..589M} {402, 589}

\bibitem[\protect\citeauthoryear{{Mansfield}, {Kravtsov}  \&
  {Diemer}}{{Mansfield} et~al.}{2017}]{shellfish}
{Mansfield} P.,  {Kravtsov} A.~V.,   {Diemer} B.,  2017, \mn@doi [\apj]
  {10.3847/1538-4357/aa7047}, \href
  {https://ui.adsabs.harvard.edu/abs/2017ApJ...841...34M} {841, 34}

\bibitem[\protect\citeauthoryear{{More}, {Kravtsov}, {Dalal}  \&
  {Gottl{\"o}ber}}{{More} et~al.}{2011}]{more_11_fof}
{More} S.,  {Kravtsov} A.~V.,  {Dalal} N.,   {Gottl{\"o}ber} S.,  2011, \mn@doi
  [\apjs] {10.1088/0067-0049/195/1/4}, \href
  {http://adsabs.harvard.edu/abs/2011ApJS..195....4M} {195, 4}

\bibitem[\protect\citeauthoryear{{More}, {Diemer}  \& {Kravtsov}}{{More}
  et~al.}{2015}]{More2015}
{More} S.,  {Diemer} B.,   {Kravtsov} A.~V.,  2015, \mn@doi [\apj]
  {10.1088/0004-637X/810/1/36}, \href
  {https://ui.adsabs.harvard.edu/\#abs/2015ApJ...810...36M} {810, 36}

\bibitem[\protect\citeauthoryear{{Oman}, {Hudson}  \& {Behroozi}}{{Oman}
  et~al.}{2013}]{oman_13}
{Oman} K.~A.,  {Hudson} M.~J.,   {Behroozi} P.~S.,  2013, \mn@doi [\mnras]
  {10.1093/mnras/stt328}, \href
  {https://ui.adsabs.harvard.edu/abs/2013MNRAS.431.2307O} {431, 2307}

\bibitem[\protect\citeauthoryear{{Press} \& {Schechter}}{{Press} \&
  {Schechter}}{1974}]{Press-Schechter1974}
{Press} W.~H.,  {Schechter} P.,  1974, \mn@doi [\apj] {10.1086/152650}, \href
  {https://ui.adsabs.harvard.edu/abs/1974ApJ...187..425P} {187, 425}

\bibitem[\protect\citeauthoryear{{Reed}, {Gardner}, {Quinn}, {Stadel},
  {Fardal}, {Lake}  \& {Governato}}{{Reed} et~al.}{2003}]{reed_03}
{Reed} D.,  {Gardner} J.,  {Quinn} T.,  {Stadel} J.,  {Fardal} M.,  {Lake} G.,
   {Governato} F.,  2003, \mn@doi [\mnras] {10.1046/j.1365-2966.2003.07113.x},
  \href {http://adsabs.harvard.edu/abs/2003MNRAS.346..565R} {346, 565}

\bibitem[\protect\citeauthoryear{{Shapiro}, {Iliev}  \& {Raga}}{{Shapiro}
  et~al.}{1999}]{Shapiroetal99}
{Shapiro} P.~R.,  {Iliev} I.~T.,   {Raga} A.~C.,  1999, \mn@doi [\mnras]
  {10.1046/j.1365-8711.1999.02609.x}, \href
  {https://ui.adsabs.harvard.edu/abs/1999MNRAS.307..203S} {307, 203}

\bibitem[\protect\citeauthoryear{{Sheth} \& {Tormen}}{{Sheth} \&
  {Tormen}}{2002}]{shethtormen}
{Sheth} R.~K.,  {Tormen} G.,  2002, \mn@doi [\mnras]
  {10.1046/j.1365-8711.2002.04950.x}, \href
  {https://ui.adsabs.harvard.edu/abs/2002MNRAS.329...61S} {329, 61}

\bibitem[\protect\citeauthoryear{{Shi}}{{Shi}}{2016}]{Shi2016}
{Shi} X.,  2016, \mn@doi [\mnras] {10.1093/mnras/stw925}, \href
  {https://ui.adsabs.harvard.edu/abs/2016MNRAS.459.3711S} {459, 3711}

\bibitem[\protect\citeauthoryear{{Springel}}{{Springel}}{2005}]{Springel2005}
{Springel} V.,  2005, \mn@doi [\mnras] {10.1111/j.1365-2966.2005.09655.x},
  \href {http://adsabs.harvard.edu/abs/2005MNRAS.364.1105S} {364, 1105}

\bibitem[\protect\citeauthoryear{{Springel}}{{Springel}}{2015}]{ngenic}
{Springel} V.,  2015, {N-GenIC: Cosmological structure initial conditions},
  Astrophysics Source Code Library, record ascl:1502.003 (\mn@eprint {ascl}
  {1502.003})

\bibitem[\protect\citeauthoryear{{Sugiura}, {Nishimichi}, {Rasera}  \&
  {Taruya}}{{Sugiura} et~al.}{2020}]{sugiura_20}
{Sugiura} H.,  {Nishimichi} T.,  {Rasera} Y.,   {Taruya} A.,  2020, \mn@doi
  [\mnras] {10.1093/mnras/staa413}, \href
  {https://ui.adsabs.harvard.edu/abs/2020MNRAS.493.2765S} {493, 2765}

\bibitem[\protect\citeauthoryear{{Tinker}, {Kravtsov}, {Klypin}, {Abazajian},
  {Warren}, {Yepes}, {Gottl{\"o}ber}  \& {Holz}}{{Tinker}
  et~al.}{2008}]{tinker_08}
{Tinker} J.,  {Kravtsov} A.~V.,  {Klypin} A.,  {Abazajian} K.,  {Warren} M.,
  {Yepes} G.,  {Gottl{\"o}ber} S.,   {Holz} D.~E.,  2008, \mn@doi [\apj]
  {10.1086/591439}, \href {http://adsabs.harvard.edu/abs/2008ApJ...688..709T}
  {688, 709}

\bibitem[\protect\citeauthoryear{{Tomooka}, {Rozo}, {Wagoner}, {Aung}, {Nagai}
  \& {Safonova}}{{Tomooka} et~al.}{2020}]{Tomooka2020}
{Tomooka} P.,  {Rozo} E.,  {Wagoner} E.~L.,  {Aung} H.,  {Nagai} D.,
  {Safonova} S.,  2020, \mn@doi [\mnras] {10.1093/mnras/staa2841}, \href
  {https://ui.adsabs.harvard.edu/abs/2020MNRAS.499.1291T} {499, 1291}

\bibitem[\protect\citeauthoryear{{Watson}, {Iliev}, {D'Aloisio}, {Knebe},
  {Shapiro}  \& {Yepes}}{{Watson} et~al.}{2013}]{watsonetal13}
{Watson} W.~A.,  {Iliev} I.~T.,  {D'Aloisio} A.,  {Knebe} A.,  {Shapiro} P.~R.,
    {Yepes} G.,  2013, \mn@doi [\mnras] {10.1093/mnras/stt791}, \href
  {https://ui.adsabs.harvard.edu/abs/2013MNRAS.433.1230W} {433, 1230}

\bibitem[\protect\citeauthoryear{{Zu} \& {Weinberg}}{{Zu} \&
  {Weinberg}}{2013}]{ZW2013}
{Zu} Y.,  {Weinberg} D.~H.,  2013, \mn@doi [\mnras] {10.1093/mnras/stt411},
  \href {https://ui.adsabs.harvard.edu/abs/2013MNRAS.431.3319Z} {431, 3319}

\makeatother
\end{thebibliography}
